\documentclass[conference]{IEEEtran}
\usepackage{bbm}
\usepackage{amssymb}
\usepackage{flushend}
\usepackage{algorithmic}
\usepackage{array}

\usepackage{pdfpages}
\usepackage{epstopdf}
\usepackage[cmex10]{amsmath}
\usepackage{graphicx}
\graphicspath{ {images/} }
\usepackage{algorithmic}

\usepackage{array}

\usepackage{pdfpages}

\usepackage{fancyhdr}
\usepackage{kantlipsum}
\renewcommand{\headrulewidth}{0pt}
\renewcommand{\footrulewidth}{0pt}

\fancypagestyle{plain}{
  \fancyhf{}
  \renewcommand{\footrulewidth}{0pt} 
  \renewcommand{\headrulewidth}{0pt}
}
\usepackage{eso-pic}

\let\svthefootnote\thefootnote

\begin{document}

\title{The Power of $d$ Choices in Scheduling for Data Centers with Heterogeneous Servers}

\author{\IEEEauthorblockN{Amir Moaddeli}
\IEEEauthorblockA{\textit{Department of Electrical Engineering} \\
\textit{Azad University of Shiraz}}
\and
\IEEEauthorblockN{Iman Nabati Ahmadi}
\IEEEauthorblockA{\textit{Department of Mechanical Engineering} \\
\textit{Yasouj University}}
\and
\IEEEauthorblockN{Negin Abhar}
\IEEEauthorblockA{\textit{Department of English Language} \\
\textit{University of Tehran}}

}

\maketitle

\thispagestyle{plain}
\pagestyle{plain}


\begin{abstract}
MapReduce framework is the de facto in big data and its applications, in which the big data-set is split into small data chunks and the data chunks are replicated on different servers, three servers by default, among thousands of servers. The heterogeneous server structure of the system, where the pair task-server determines the task processing speed, makes the scheduling much harder than scheduling for systems with homogeneous servers. Throughput optimality of the system on one hand and delay optimality on the other hand creates a dilemma for assigning tasks to servers. Unless many research has been done on both heuristic and theoretical algorithms for scheduling for such a system with multi-level data locality, the optimal scheduling algorithm is still an open problem. The JSQ-MaxWeight and Balanced-Pandas algorithms are the states of the arts algorithms with theoretical guarantees on throughput and delay optimality for systems with two and three levels of data locality. However, the scheduling complexity of these two algorithms are way too much. Hence, we use the power of $d$ choices algorithm combined with the Balanced-Pandas algorithm and the JSQ-MaxWeight algorithm, and compare the complexity of the simple algorithms and the power of $d$ choices versions of them. We will further show that the Balanced-Pandas algorithm combined with the power of the $d$ choices, Balanced-Pandas-Pod, not only performs better than simple Balanced-Pandas, but also is less sensitive to the parameter $d$ than the combination of the JSQ-MaxWeight algorithm and the power of the $d$ choices, JSQ-MaxWeight-Pod. In fact in our extensive simulation results, the Balanced-Pandas-Pod algorithm is performing better than the simple Balanced-Pandas algorithm in low and medium loads, where data centers are usually performing at, and performs almost the same as the Balanced-Pandas algorithm at high loads. The reason is that the combination of the two algorithms enhances the data locality of tasks and prioritize local and rack local service to remote service more than the Balanced-Pandas algorithm. Note that the load balancing complexity of Balanced-Pandas and JSQ-MaxWeight algorithms are $O(M)$, where $M$ is the number of servers in the system which is in the order of thousands servers, whereas the complexity of Balanced-Pandas-Pod and JSQ-MaxWeight-Pod are $O(1)$, that makes the central scheduler faster and saves energy. Moreover, as mentioned, our proposed Balanced-Pandas-Pod has better performance than the existing algorithms that makes it superior to the state-of-the-art algorithms.
\end{abstract}

\IEEEpeerreviewmaketitle

\addtocounter{footnote}{-1}\let\thefootnote\svthefootnote

\let\thefootnote\relax\footnote{The copyright notice is: 978-1-5386-4304-4/18/\$31.00 \copyright 2018 IEEE}

\section{Introduction}
\label{introduction}

The optimum task scheduling problem in data centers is a special case of affinity scheduling problem, which is an open problem. The affinity scheduling problem is to dynamically assign multi-type tasks to multi-skilled servers in an optimum way to minimize a cost function , for example the mean task completion time. The completion time of a task consists of waiting and service times, which is an important factor for user satisfaction of the service they receive from the system.

Scheduling for data centers are mainly divided into two categories, the first one is for the case of homogeneous servers where all servers are capable of giving service to all types of tasks and the service time has the same distribution for a task in all servers. This first category of scheduling has been studied in details and the results on this kind of scheduling are summarized as follows. Consider the existence of one queue per server, where the arriving tasks to the server that cannot get service immediately due to the fact that all servers are busy are queued at these queues. The join-the-shortest-queue (JSQ) is the optimum algorithm in the homogeneous servers scenario that minimizes the mean completion time for all tasks. The JSQ algorithm routes an incoming task to the server that has the minimum queue length in the whole system and schedules an idle server to process a task that is queued in the queue associated to the server. However, assuming an ordinary data center with thousands of servers, it is a costly function for the centralized scheduler to constantly search the queue with the minimum queue length and routes the incoming task to that queue. Moreover, the messaging costs between the servers and the centralized scheduler is proven to be costly and add a considerable overhead on the networking system of data centers which makes the JSQ algorithm complicated to be used in data centers \cite{ying2017power}. Therefore, a simplified version of JSQ was proposed by Mitzenmacher \cite{mitzenmacher1996load, mitzenmacher2001power, richa2001power, byers2004geometric, lumetta2006using} which later was called the power-of-d-choices (Pod) algorithm. In the Pod algorithm, at the arrival of a task, the central scheduler samples $d$ servers uniformly at random and routed the task to the queue with the minimum queue length among those $d$ queues. This algorithm is proven to be throughput optimal for a system with homogeneous servers when $d \geq 2$. Even though the Pod is not Heavy-traffic (delay) optimal, it has a comparable performance with the JSQ algorithm where the proof follows the classical balls and bins proof. Suppose that $n$ balls are going to be placed in $n$ bins. For each ball, if we sample all the bins and place it in the one with the minimum number of balls (ties are broken at random), the load on each bin would be one ball, but as mentioned, the sampling process is costly in data center applications. On the other end, if each ball is placed at a bin chosen uniformly at random, the maximum load in the $n$ bins is approximately $\frac{\log n}{\log\log n}$ with high probability which is in the order of $\log n$ balls. Instead, if two bins are chosen uniformly at random and the ball is placed in the bin with fewer balls, the maximum load is decreased to $\frac{\log\log n}{\log d} + O(1)$ which is in the order of $\log\log n$. Hence, sampling two queues and placing the ball in the lower loaded bin instead of placing the ball in a random bin decreases the maximum load of the system to $\log\log n$ from $\log n$. Note that the overhead on messaging communications for the JSQ algorithm is equal to the number of total tasks multiplied by the number of servers since the central scheduler needs the queue length of all servers for each task, but on the other hand, the messaging overhead for the Po2 algorithm is two times the number of tasks which is way lower than the JSQ overhead. Even though the Po2 algorithm is not heavy-traffic optimal (delay optimal), it has an extremely lower overhead which made it suitable for use in data centers.

The second category of scheduling algorithms is for the case of heterogeneous servers, where all servers are capable of giving service to all types of tasks, but with different service distributions. More specifically, one task type can have fast service rate in servers $1, 2$, and $3$, while another task type has slow service rate in servers $1, 2$, and $3$ and has fast service rate in servers $6, 8$, and $9$. Hence, the heterogeneity of the servers is different from the perspective of different task types which makes the scheduling problem for such a system much harder than the homogeneous scenario, where the general case is the affinity scheduling problem which is an open problem. Such a system model fits into the MapReduce applications which are widely being used in Hadoop \cite{hadoop}, Google's MapReduce \cite{dean2008mapreduce}, grid-computing \cite{isard2009quincy}, \cite{chehardeh2016remote}, \cite{almalki2015capacitor}, \cite{chehardeh2018systematic}, and Dryad \cite{isard2007dryad}. MapReduce is the de facto standard in processing big data, where the big data-set is divided into typical small chunks of data of size 128 MB and each chunk of data is replicated on a number of servers, where the default in Hadoop is three servers. The replication of data chunk is for fault tolerance of the data center and availability of data chunks. The processing of the big data-set is now divided into tasks (also called map tasks), where each task corresponds to processing a data chunk. After all map tasks are processed, one or multiple reduce tasks combine the results of all map tasks to produce the final result on the big data-set. Note that processing of big data-sets are either map-intensive or only consist of map task \cite{MaxWeight, BalancedPandas, yekkehkhany2017near, daghighi2017scheduling, kavousi2017affinity}, so our focus in this paper is on map task scheduling. Each map task can be processed by any of the servers, but the service rate is faster in the servers that have the data chunk associated with the task, which are referred as local servers. If a task receives service from a non-local server, the processing time is longer since the server should first fetch the data from one of the local servers, then start on the process. Unless the speed of data center networks has been increased by magnitudes, data transfer still causes delay due to congestion in the network that non-local servers can have up to 6 times slower service rate than local servers \cite{ananthanarayanan2012pacman, ananthanarayanan2011scarlett, zaharia2010delay}. Hence, based on the location of the local servers of tasks, the map tasks are divided into different types of task where the heterogeneity of the servers is different from the perspective of these task types. The system model is discussed in more details in Section \ref{systemmodel}, where a data center with three levels of data locality (three processing rates at three groups of servers for each map task) is studied, where the data locality levels appear due to the rack structure of data centers. Due to the heterogeneous servers, the JSQ algorithm or the joint-the-shortest-local-queue are not optimum in the sense that they are not even throughput optimal. Recent works by Xie et al. on the Balanced-Pandas algorithm \cite{BalancedPandas, yekkehkhany2017near} and Wang et al. on JSQ-MaxWeight \cite{MaxWeight} focus on the routing and scheduling policies of load balancing algorithms for data center with two or three levels of data locality. It has been shown that both the Balanced-Pandas and JSQ-MaxWeight algorithms are throughput optimal, but the Balanced-Pandas algorithm is heavy-traffic optimal in the whole capacity region of the data center, while the JSQ-MaxWeight algorithm is heavy-traffic optimal in a specific traffic load. Hence, the Balanced-Pandas algorithm is the state-of-the-art data center load balancing algorithm. However, both the JSQ-MaxWeigth and Balanced-Pandas algorithms have the load balancing complexity similar to what is mentioned for the JSQ algorithm in the homogeneous scenario. Even though the Balanced-Pandas algorithm is heavy-traffic optimal, at the arrival of a task, the central scheduler should find the server with the minimum weighted workload among all servers to route the task to that server. On the other hand, when a server becomes idle, the JSQ-MaxWeight algorithm should find the queue with the maximum length in the system, which is used in scheduling of the idle server to a task. In both cases, the messaging overhead and the computing capacity needed to find the server with the minimum weighted workload or to find the server that has the maximum queue length can be way too much for a data center with thousands of servers.

In all the above mentioned algorithms, the processing rates of servers for different task types are assumed to be known. However, the processing rates of different task types on servers are usually unknown and change with time. In a recent work by Yekkehkhany and Nagi \cite{yekkehkhany2019blind}, they proposed an exploration-exploitation-based algorithm, the Blind GB-PANDAS algorithm, that addresses this issue. This method is out of the scope of this paper, but is a promising future work to be combined with our proposed algorithm. As a complementary method, reinforcement learning tools can be utilized to model the data center behavior to estimate the processing rates of task types on servers as used in scenarios where both manned and unmanned vehicles coexist \cite{musavi2016unmanned}, \cite{musavi20183d}, \cite{lee2013counter}, \cite{jaakkola1995reinforcement}, \cite{musavi2016game}, and \cite{yildiz2013predicting}.

In this work, we propose the combination of the Balanced-Pandas and the Power-of-d-choices algorithms (Balanced-Pandas + Pod or Balanced-Pandas-Pod) and the combination of the JSQ-MaxWeight and the Power-of-d-choices algorithms (JSQ-MaxWeight + Pod or JSQ-MaxWeight-Pod) for a typical data center with three levels of data locality. Note that in a system like Hadoop that has more than one server with the highest rate for a specific task, Po2 wastes the resources and does not even stabilize the system in its capacity region. In Hadoop, the default is to have three local servers for each task, so the combination of the Balanced-Pandas or JSQ-MaxWeight algorithms with Pod is useful for $d \geq 3$. We run extensive simulations and show that the combination of Pod with both the Balanced-Pandas and JSQ-MaxWeight algorithms are throughput optimal, but the Balanced-Pandas-Pod performs better than JSQ-MaxWeight-Pod up to tenfold at high loads. It is also notable in our simulation results that if there are $M$ servers in the data center, even though the Balanced-Pandas-Pod has $O(1)$ scheduling complexity versus the Balanced-Pandas having $O(M)$ scheduling complexity, the Balanced-Pandas-Pod performs better than the Balanced-Pandas at low and medium loads, where the data centers are usually performing at, and at high loads, both algorithms are having almost the same delay performance. Hence, we conclude that our novel proposed algorithm, the Balanced-Pandas-Pod algorithm, beats the state-of-the-art Balanced-Pandas algorithm, also our algorithm has less scheduling complexity.

\section{Related Work}

As of previous work done on load balancing for data centers, both heuristic and theoretical algorithms have been proposed. Among the heuristic algorithms, the de facto standard in Hadoop is the delay Fair scheduler \cite{hadoop}, but simple facts like the optimum delay which depends on the load of the system are not studied. For more details on different widely used heuristic algorithms, see \cite{jin2011bar, zaharia2008improving, he2011matchmaking, zaharia2010delay, polo2011resource, ibrahim2012maestro, isard2009quincy, white2010hadoop}. As mentioned in the Introduction, the data center load balancing is an application of affinity scheduling problem. The classic studies on affinity scheduling include Fluid Model Planning by Harrison and Lopez \cite{harrison1998heavy, harrison1999heavy}, resource pooling by Bell and Williams \cite{bell2005dynamic, bell2001dynamic}, and the generalized c$\mu$-rule algorithm which is based on MaxWeight scheduling by Stolyar and Mandelbaum \cite{mandelbaum2004scheduling, stolyar2004maxweight}. A mutual problem with all these algorithms is that they all consider having a separate queue for each task type. In the data center application, the number of task types is so large, in the cubic order of number of tasks, that makes it impossible to have separate queues for task types. The reason that the number of task types is in the cubic order of the number of servers is that, having $M$ servers, each task is saved on three servers in Hadoop, so there can be ${M \choose 3} = O(M^3)$ number of task types. Another problem with Fluid Model Planning is that it requires the knowledge of arrival rate of all task types which is not a reasonable assumption since the load on data centers varies a lot and cannot be estimated precisely. Moreover, the generalized c$\mu$-rule does not minimize the mean task completion time, but minimizes a cost function which is not of interest in data center application. All these problems with these classic algorithms make them impractical for data center load balancing. As mentioned earlier, the recent works by Xie et al. on Balanced-Pandas \cite{BalancedPandas}, Wang et al. on JSQ-MaxWeight \cite{MaxWeight}, and Yekkehkhany et al. on GB-PANDAS \cite{yekkehkhany2017gb} solves the problems of not having the arrival rates of all task types and not having per task type queue, but their scheduling complexity is still high, where in each case the central scheduler needs to do $O(M)$ calculation for routing a task or scheduling an idle server, where in our proposed Balanced-Pandas-Pod and JSQ-MaxWeight-Pod algorithms, the routing and scheduling complexities are $O(1)$, which makes load balancing much faster and saves more energy, besides having better performance than the state-of-the-art algorithms in low and medium loads, where data centers are normally performing at, and having almost same performance at high loads as the optimum Balanced-Pandas.

\section{{SYSTEM MODEL}}
\label{systemmodel}
\begin{figure*}
	\includegraphics[width=\linewidth]{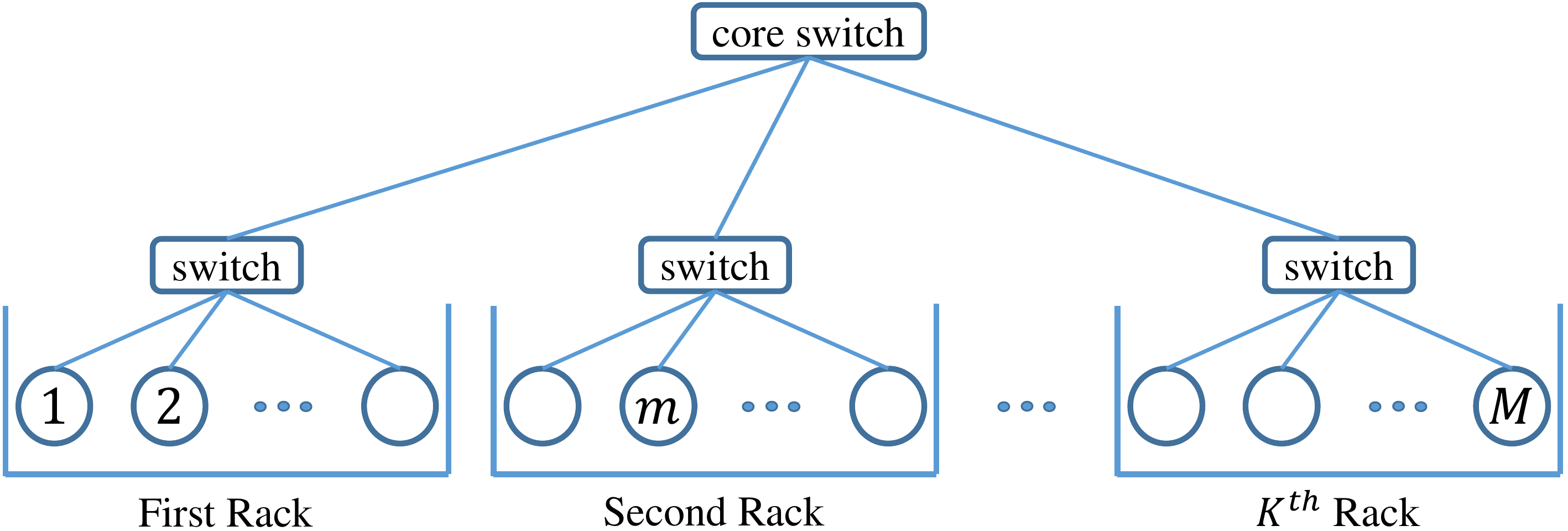}
	\caption{The typical data center structure with racks, top of the rack switches and a core switch.}
	\label{serversmodel}
\end{figure*}

We study a discrete time model where time is indexed by $t \in \{0, 1, 2, 3, \cdots\}$. There are $M$ servers in the system, where they are denoted by $m \in \mathcal{M} = \{1, 2, 3, \cdots, M\}$. The $M$ servers are not isolated from each other, but they are connected to each other by switches on top of the racks and a core switch as shown in figure \ref{serversmodel}.
The servers that are directly connected to each other through a top of the rack switch form a rack. All racks are of the same size, and a rack is indexed by $k \in \mathcal{K} = \{1, 2, \cdots, K\}$. By a little bit misuse of notation, let the outcome of the function $K(m)$ be the index of the rack that the $m$-th server belongs to, which ranges from $1$ to $K$.

As mentioned in the introduction, the data chunk associated to a task is saved on three servers in Hadoop \cite{hadoop} that determines the type of a task type. Hence, the type of a task is denoted by $\bar{L} = (m_1, m_2, m_3)$, where $m_1, m_2,$ and $m_3$ are the three servers that have the corresponding data chunk and are called local servers. Note that the set of all task types is $\mathcal{L} = \{ (m_1, m_2, m_3) : m_1, m_2, m_3 \in \mathcal{M} \text{ and } m_1 < m_2 < m_3 \}$. The servers in the set $\bar{L}_k = \{m \notin \bar{L}: K(m) = K(m_1) \text{ or } K(m) = K(m_2) \text{ or } K(m) = K(m_3) \}$ are called rack-local servers for the task of type $\bar{L}$ since they do not have the corresponding data chunk, but another server in their rack has it, so they can fetch the data through a single switch, which is the top of the rack switch. Unless there is a delay in fetching the data from another server in the same rack, it is on average shorter than fetching data from another server in another rack. The servers in the set $\bar{L}_r = \{m \in \mathcal{M}: m \notin \bar{L} \text{ and } m \notin \bar{L}_k \}$ are called remote servers, where fetching data in a remote server takes much longer time compared to accessing the data locally or rack-locally. A task can practically receive service from any of the servers in the system, but the processing rate in a local server is fastest, then rack-local, and slowest on a remote server. Assume that the processing duration of a task on a local, rack-local, and remote server has Geometric distribution with parameters $\alpha, \beta,$ and $\gamma$, respectively, where $\alpha > \beta > \gamma$. Then the mean processing time on a local, rack-local, and remote server is $\frac{1}{\alpha}, \frac{1}{\beta},$ and $\frac{1}{\gamma}$, respectively, where $\frac{1}{\alpha} < \frac{1}{\beta} < \frac{1}{\gamma}$. On the other hand, the number of task arrivals of type $\bar{L}$ at time slot $t$ is denoted by $A_{\bar{L}}(t)$, where $\mathbb{E}[A_{\bar{L}}(t)] = \lambda_{\bar{L}}$ and the task arrivals are independent both across different task types and different time slots. We denote the arrival rates of all task types by the vector $\boldsymbol{\lambda} = \{\lambda_{\bar{L}} : \bar{L} \in \mathcal{L} \}$. Given the assumed model for the arrival and processing of processes of the task types, the capacity of the system is given in the following subsection.

\subsection{Capacity Region}
Assuming that a sever can process with rate one, the system can be stabilized if the local, rack-local, and remote load on each server is strictly less than one. Consider the decomposition of the vector $\boldsymbol{\lambda}$ as follows. The arrival rate of task of type $\bar{L}$, $\lambda_{\bar{L}}$, is decomposed to $\lambda_{\bar{L}, m}$ for $m \in \mathcal{M}$, where $\lambda_{\bar{L}, m}$ is the arrival rate of task of type $\bar{L}$ that is assigned to server $m$. Then, the capacity region of the system is characterized as follows:
\[
\begin{aligned}
\Lambda = \Big \{ & \boldsymbol{\lambda}: \forall \bar{L}, \  \exists \lambda_{\bar{L}, m} \geq 0, \forall m \in \mathcal{M}, \text{ s.t.} \\
& \lambda_{\bar{L}} = \sum_{m = 1}^M \lambda_{\bar{L}, m}, \\
& \sum_{\bar{L}: m \in \bar{L}} \frac{\lambda_{\bar{L}, m}}{\alpha} + \sum_{\bar{L}: m \in \bar{L}_k} \frac{\lambda_{\bar{L}, m}}{\beta} + \sum_{\bar{L}: m \in \bar{L}_r} \frac{\lambda_{\bar{L}, m}}{\gamma} < 1. \Big \} 
\end{aligned}
\]

\section{{SCHEDULING ALGORITHMS}}
\label{schedulingalgorithms}
Assuming the system model described in Section \ref{systemmodel}, the Balanced-Pandas \cite{BalancedPandas} and JSQ-MaxWeight \cite{MaxWeight} algorithms are described in Subsections \ref{sub1} and \ref{sub2}, then our proposed algorithms, the Balanced-Pandas-Pod and JSQ-MaxWeight-Pod algorithms, are proposed in Subsections \ref{sub3} and \ref{sub4}. Note that the queueing structure in the Balanced-Pandas(-Pod) algorithms is different from the one in the JSQ-MaxWeight(-Pod) algorithms, which is illustrated in the corresponding subsections. Each algorithm consists of two parts, routing and scheduling, where routing is for queueing a new arrived task to a queue in the system, and scheduling is for assigning an idle server to a task in the system.

\subsection{The Balanced-Pandas Algorithm}
\label{sub1}
Consider the existence of three queues per each server. The three queues of the $m$-th server are denoted by $Q_m^1, Q_m^2,$ and $Q_m^3$ that keep the local, rack-local, and remote tasks of the server, respectively. 
The length of the local, rack-local, and remote queues of the $m$-th server at time slot $t$ are denoted by $Q_m^1(t), Q_m^2(t),$ and $Q_m^3(t)$, respectively. The workload on the $m$-th server is defined as the mean processing time of all three types of tasks that are queued in the three queues of the server. Denoting the workload of the $m$-th server at time slot $t$ by $W_m(t)$, we have $W_m(t) = \frac{Q_m^1(t)}{\alpha} + \frac{Q_m^2(t)}{\beta} + \frac{Q_m^3(t)}{\gamma}$. Then, the routing and scheduling policies of the Balanced-Pandas algorithm are as follows. An incoming task of type $\bar{L}$ is \textbf{routed} to the server with the minimum weighted workload, i.e. it is queued at the corresponding sub-queue of a server in the set $\underset{m \in \mathcal{M}}{ArgMin} \ \bigg \{ \frac{W_m(t)}{\alpha} I_{\{ m\in \bar{L} \}},\frac{W_m(t)}{\beta} I_{\{ m\in \bar{L}_k \}}, \frac{W_m(t)}{\gamma} I_{\{ m\in \bar{L}_r \}}  \bigg \}$. An idle server $m$ is first \textbf{scheduled} to give service to the local tasks queued at $Q_m^l$, but if there are no tasks available at $Q_m^l$, the server is scheduled to give service to a rack-local task queued at $Q_m^k$, finally if both $Q_m^l$ and $Q_m^k$ are empty, the service is given to remote tasks queued at $Q_m^r$. If all the sub-queues of the idle server are empty, the server remains idle till a new task is routed to one of its sub-queues. Note that it is proven that the Balanced-Pandas algorithm is both throughput and heavy-traffic optimal for a system with three levels of data locality.

\subsection{The JSQ-MaxWeight Algorithm}
\label{sub2}

Consider one queue per server denoted by $Q_m$ for the $m$-th server. The queue length of the $m$-th server at time slot $t$ is denoted by $Q_m(t)$. An incoming task of type $\bar{L}$ is routed to the shortest local queue, and an idle server is scheduled to give service to a task of one of the servers in the set $\underset{n \in \mathcal{M}}{ArgMax} \ \{ \alpha Q_n(t) I_{\{ n = m \}}, \beta Q_n(t) I_{\{ K(n) = K(m) \}},$ $
\gamma Q_n(t) I_{\{ K(n) \neq K(m) \}} \}$, where $K(m)$ is the set of servers that are in the same rack of server $m$. Note that the JSQ-MAxWeight algorithm is throughput optimal, but it is not heavy-traffic optimal in all traffic scenario.

\subsection{The Balanced-Pandas-Pod Algorithm}
\label{sub3}

Consider the same queueing structure as the one for Balanced-Pandas algorithm. The routing of Balanced-Pandas algorithm is computationally expensive since for every incoming task the minimum weighted workload should be calculated among the $M$ servers. Instead, in the Balanced-Pandas-Power-of-d-choices algorithm (Balanced-Pandas-Pod), an incoming task is routed to the server with minimum weighted workload among the three local servers and another $d$ servers that are chosen uniformly at random from the rack-local and remote servers. The scheduling policy of the Balanced-Pandas-Pod algorithm is the same as the Balanced-Pandas algorithm. When $d = M - 3$, the Balanced-Pandas-Pod algorithm is the same as Balanced-Pandas, but our simulation results shows that for $d << M$, not only the routing computation is much cheaper, but also the performance is better for medium and even high loads. We expect that for very high loads too close to the capacity boundary the Balanced-Pandas-Pod does not perform as good as Balanced-Pandas, but in those loads the mean task completion time is so high that it is not of interest. Note that the computational complexity of the Balanced-Pandas-Pod is $\frac{d + 3}{M}$ of the Balanced-Pandas algorithm, where for a typical data center with $500$ servers and $d = 8$, it is $2.2\%$ which is way lower with even better performance.

\subsection{The JSQ-MaxWeight-Pod Algorithm}
\label{sub4}

Consider the same queueing structure as the JSQ-MaxWeight algorithm. The routing policy of the JSQ-MaxWeight-Power-of-d-choices algorithm (JSQ-MaxWeight-Pod) is the same as the JSQ-MaxWeight-Pod, but their scheduling policies are different. Similar to the Balanced-Pandas-Pod algorithm, a subset of servers are considered for finding the queue with the maximum weight instead of checking all servers for an idle server. In other words, when a server becomes idle, its queue length is multiplied with $\alpha$ and the queues of another $d'$ servers are multiplied by $\beta$ and $\gamma$, then the service is given to a task in the queue with the maximum weight. Our extensive simulation results shows that the JSQ-MaxWeight-Pod performs poorly in both medium and high loads. Unless we chose $d' = 12$, so that $d' + 1 = 13 > d + 3 = 11$, where $d$ is the parameter for the Balanced-Pandas-Pod, the JSQ-MaxWeight-Pod performs tens of tens of times worst than the other three algorithms at high loads which suggests not to use this algorithm. We included this algorithm to emphasize on the fact that the power of the $d$ choices does not always makes the system less computationally complicated with tolerable decay of performance. If the power of the $d$ choices is not used wisely, it can cause a big decrease in the performance of the system.

\section{{SIMULATION RESULTS}}
\label{simulationresults}
\begin{figure}
	\includegraphics[width=\linewidth]{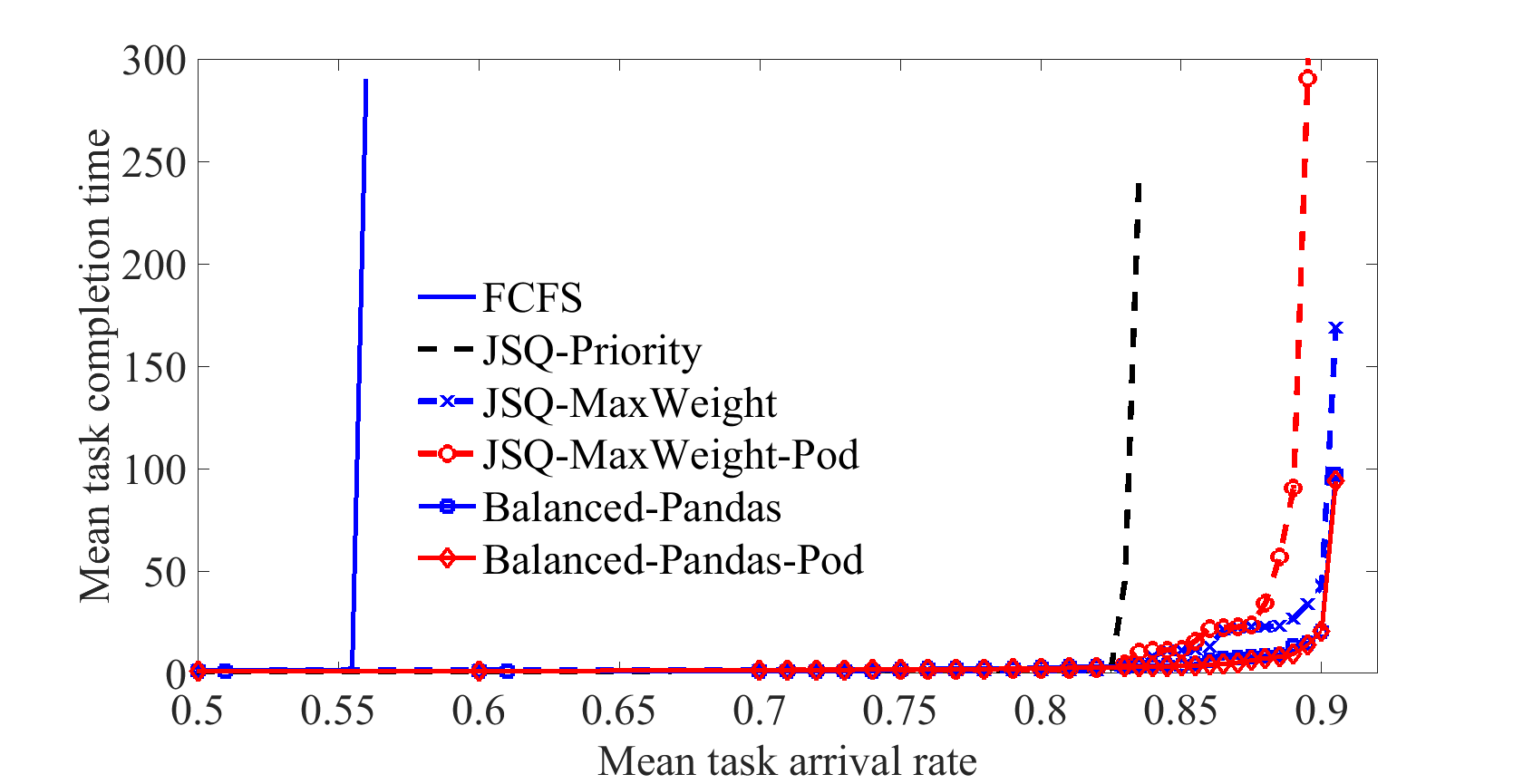}
	\caption{Performance comparison of algorithms when service time has exponential distribution.}
	\label{1exp}
\end{figure}

Consider a data center with $500$ servers where there are $10$ racks of equal size with $50$ servers each. In this section we compare six algorithms, First-Come-First-Served (FCFS), Join-the-Shortest-Queue-Priority (JSQ-Priority), JSQ-MaxWeight, JSQ-MaxWeight-Pod, Balanced-Pandas, and Balanced-Pandas-Pod in terms of the mean task completion time that they have. For the JSQ-MaxWeight-Pod, in the scheduling part we check the queue length of the idle server in addition to six rack-local servers and six remote servers uniformly at random, so the parameter of this algorithm is chosen to be $d' = 12$. For the Balanced-Pandas-Pod algorithm, in the routing part, we check the three servers, two rack-local servers and six remote servers uniformly at random, i.e. the parameter associated to this algorithm is $d = 8$. As we discussed before, the computational complexity of the Balanced-Pandas-Pod is only $2.2\%$ while it surprisingly outperforms all the other algorithms at medium and almost very high loads as it is shown later in this section. The reason why Balanced-Pandas-Pod performs this well unless it has less computational complexity is that the lower sampling of the servers is aligned with giving more priority to local servers which are faster, while doing load balancing on randomly chosen servers whenever the load is high on local servers. We considered a more realistic system that is continuous-time in our simulations instead of discrete-time. We analyzed the results both under exponential distribution for the service times and log-normal distribution which has a heavy tail. The six algorithms can also be tested on real data sets like the one in \cite{hosseini2017mobile} or \cite{alrefaie2014road}. Under exponential service time, the mean task completion time under the six algorithms is shown in figure \ref{1exp}. Obviously, the Balanced-Pandas-Pod is outperforming all the other algorithms unless we have not taken the computational complexity into account.

\begin{figure}
	\includegraphics[width=\linewidth]{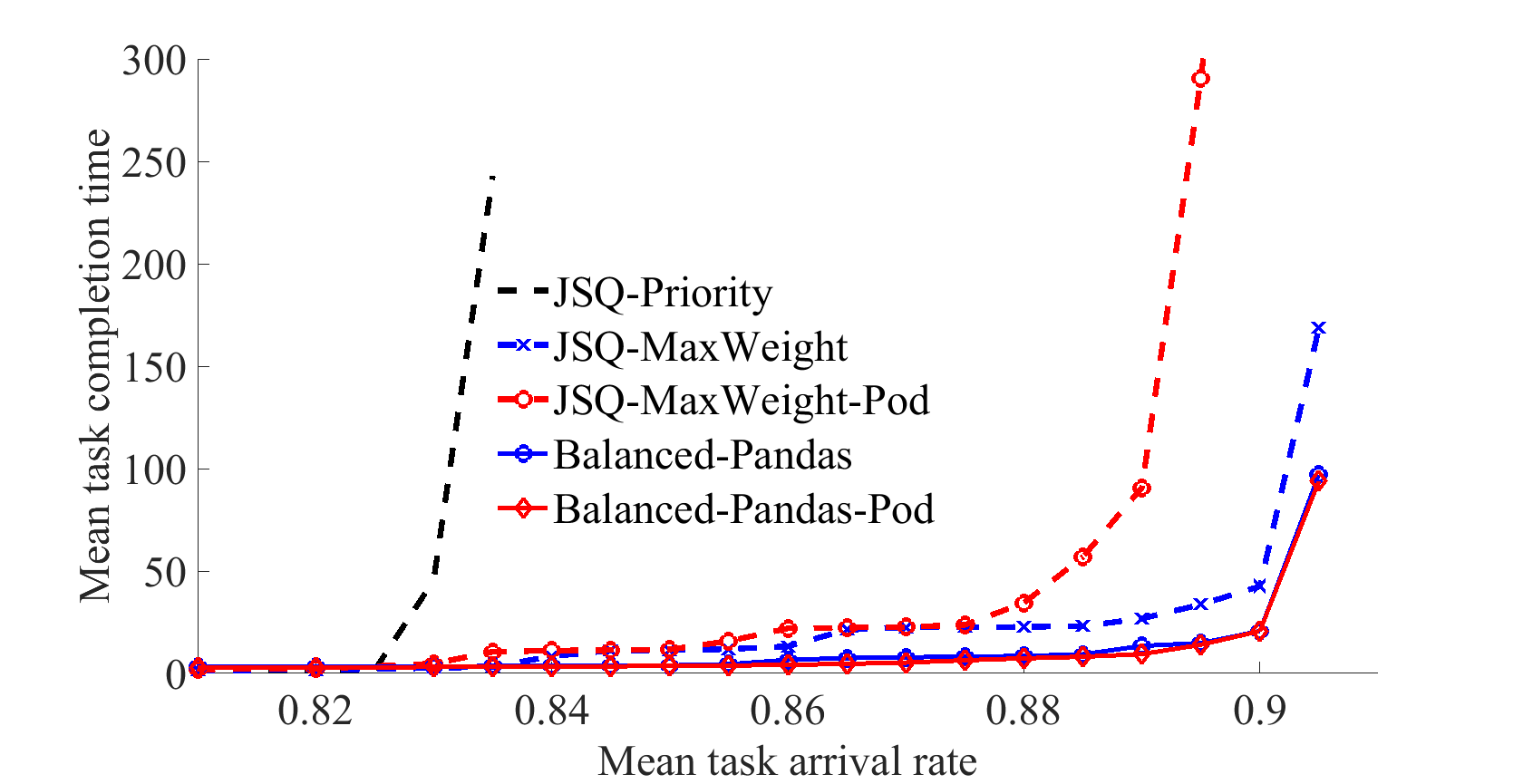}
	\caption{Performance comparison of algorithms at high loads when service time has exponential distribution.}
	\label{2exp}
\end{figure}

Zooming at high loads, figure \ref{2exp} shows the better performance of Balanced-Pandas-Pod versus the other algorithms.

\begin{figure}
	\includegraphics[width=\linewidth]{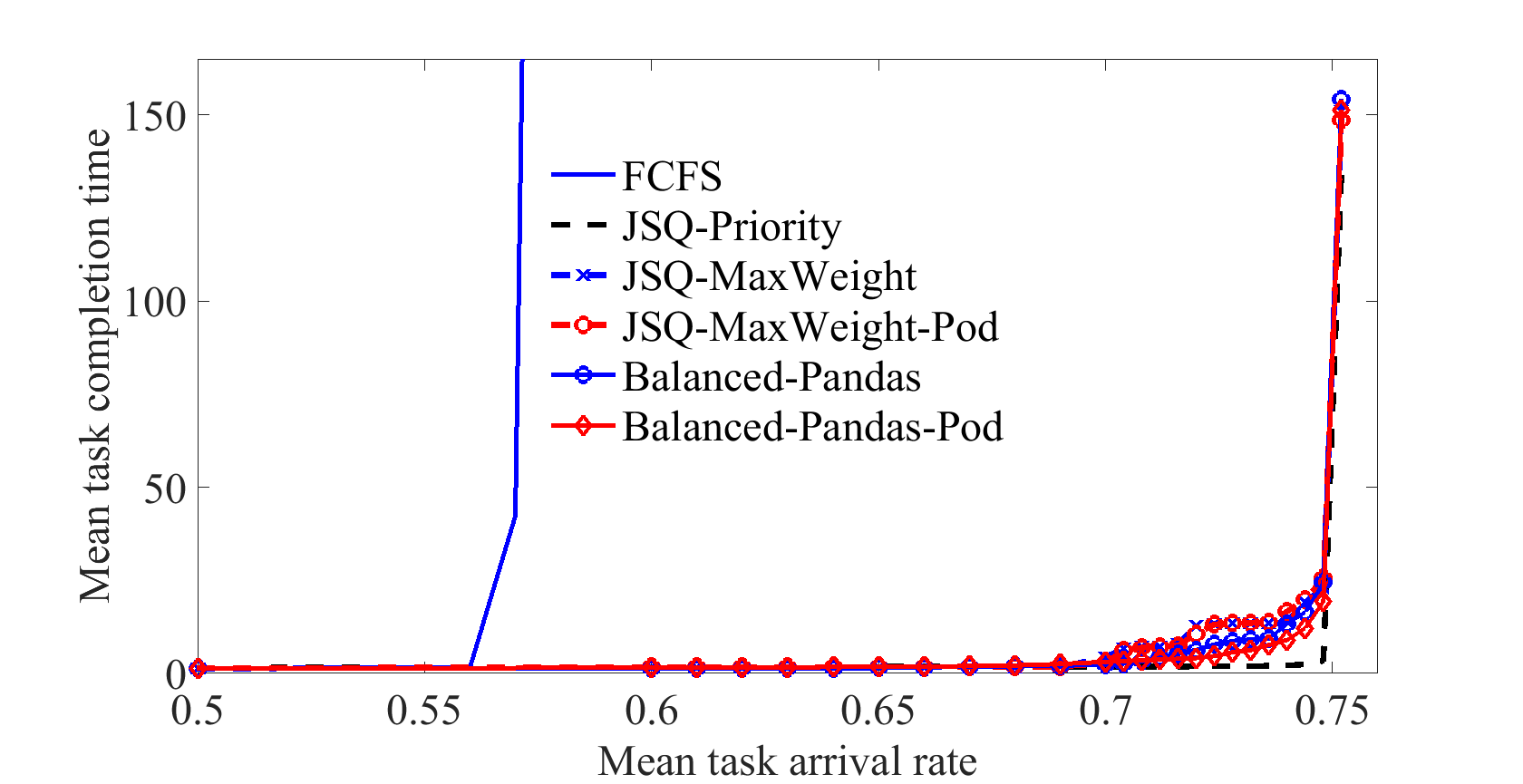}
	\caption{Performance comparison of algorithms at a specific load where Balanced-Pandas and JSQ-MaxWeight are delay optimal when service time has exponential distribution.}
	\label{2expspecialload}
\end{figure}

Finally, the performance of algorithms is compared at a specific load in which both Balanced-Pandas and JSQ-MaxWeight are delay optimal, which is shown in figure \ref{2expspecialload}

\begin{figure}
	\includegraphics[width=\linewidth]{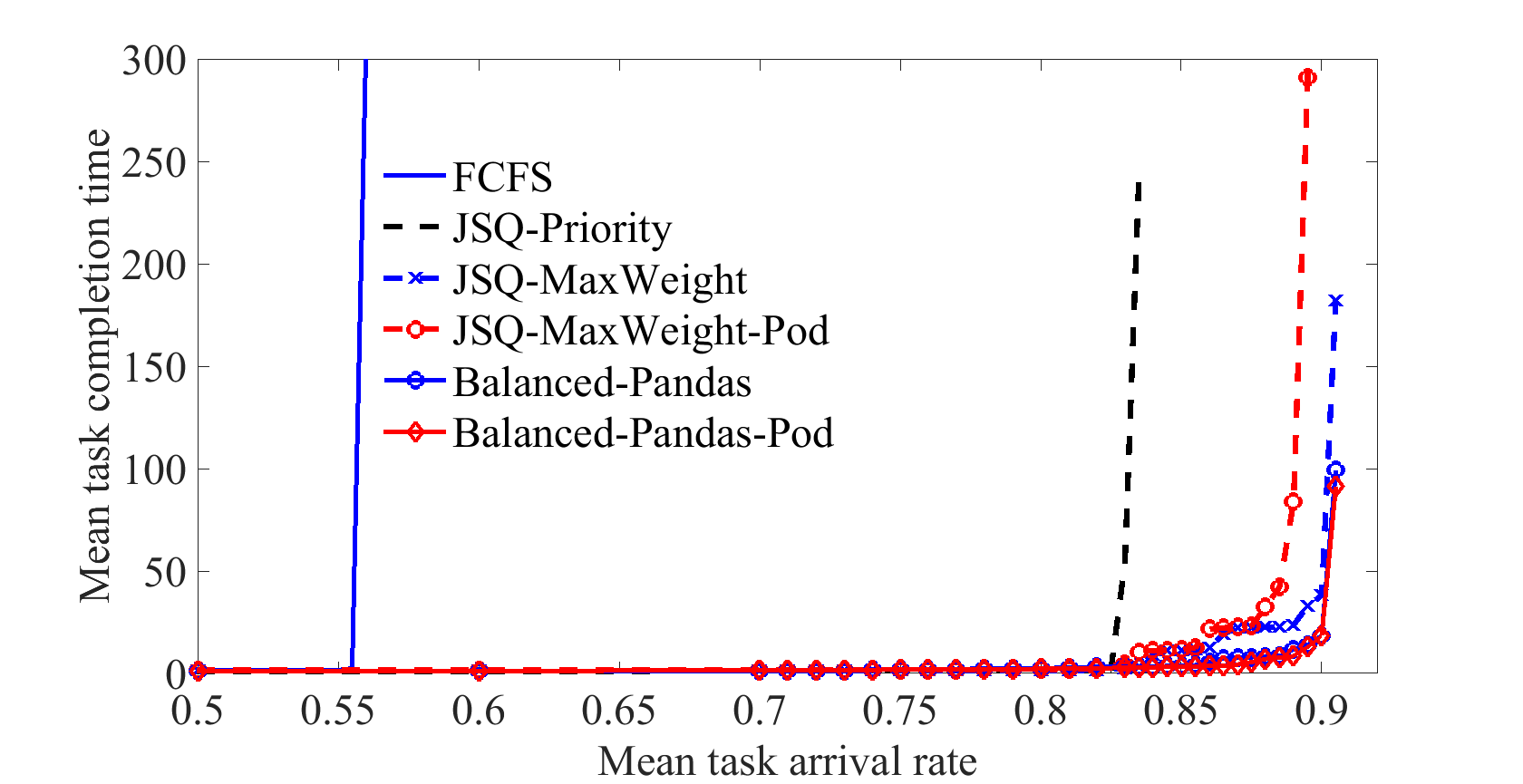}
	\caption{Performance comparison of algorithms when service time has log-normal distribution.}
	\label{3lognormal}
\end{figure}

\begin{figure}
	\includegraphics[width=\linewidth]{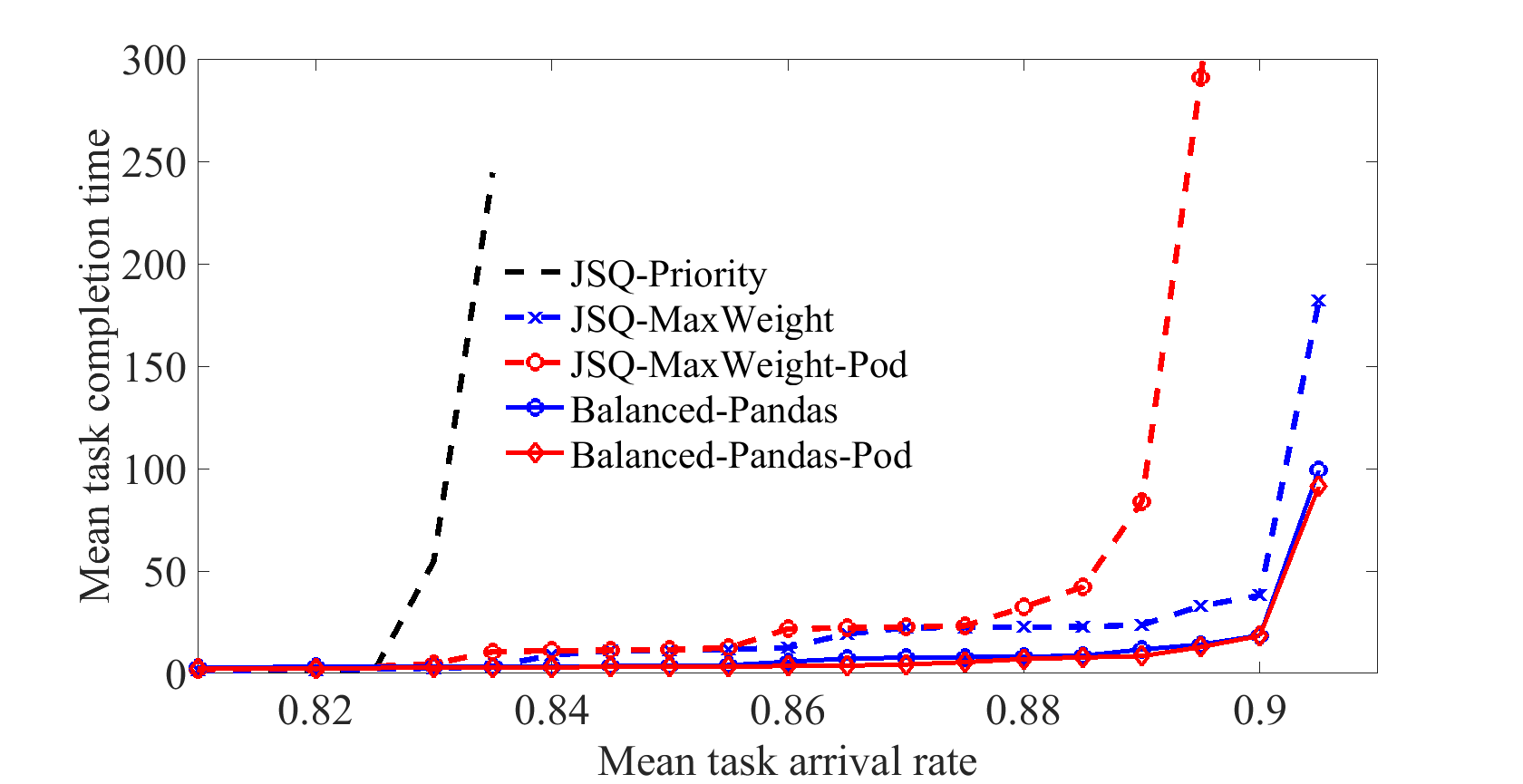}
	\caption{Performance comparison of algorithms at high loads when service time has log-normal distribution.}
	\label{33lognormal}
\end{figure}

\begin{figure}
	\includegraphics[width=\linewidth]{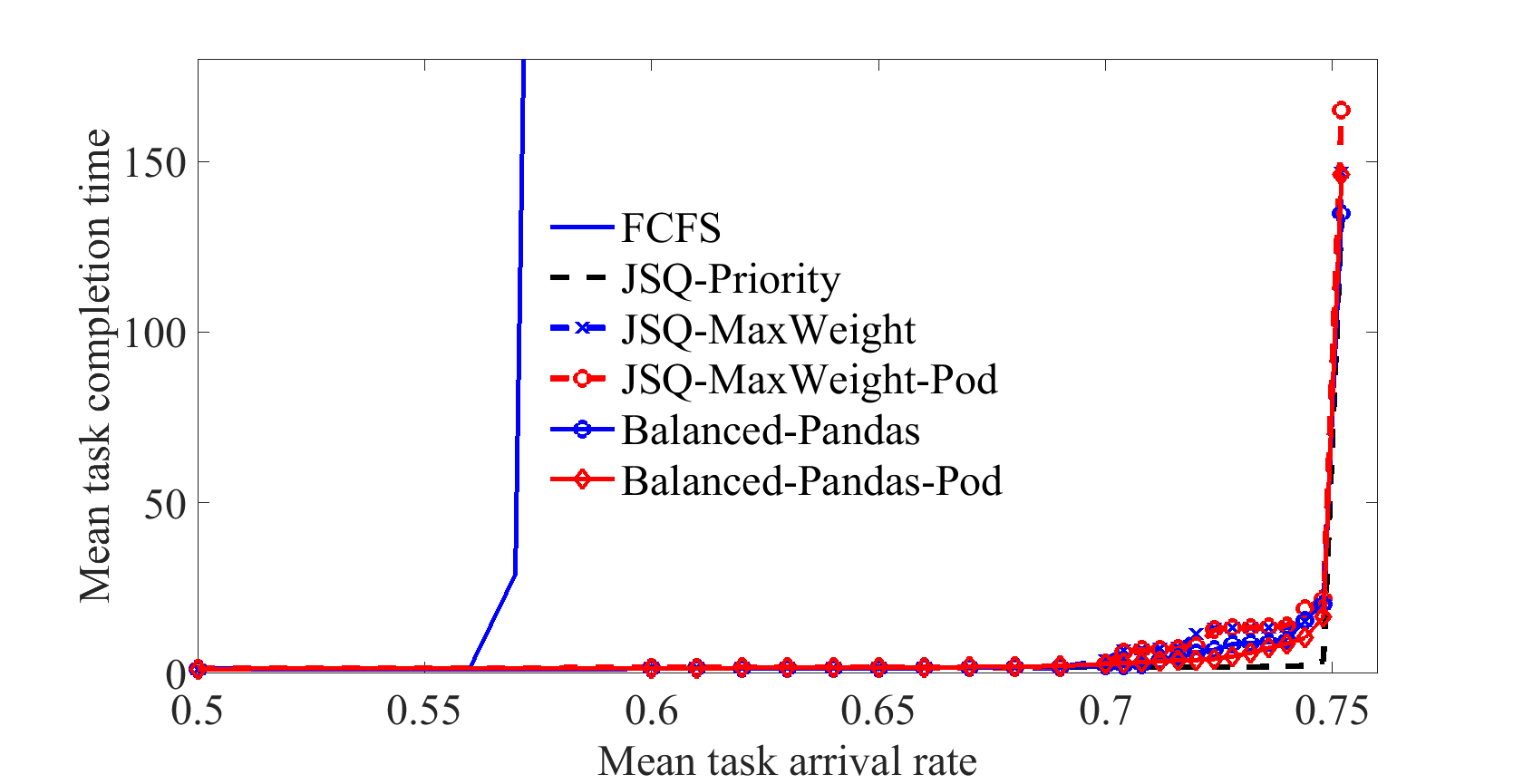}
	\caption{Performance comparison of algorithms at a specific load where Balanced-Pandas and JSQ-MaxWeight are delay optimal when service time has log-normal distribution.}
	\label{4lognormalspecialload}
\end{figure}

Similar observations are available when the service time has log-normal distribution which are shown in figures \ref{3lognormal}, \ref{33lognormal}, and \ref{4lognormalspecialload}.

\section{{CONCLUSION AND FUTURE WORK}}
\label{conclusion}

In this work, we proposed the Balanced-Pandas-Pod and JSQ-MaxWeight-Pod algorithms that have $2.2\%$ computational complexity of the Balanced-Pandas and JSQ-MaxWeight algorithms, while the Balanced-Pandas-Pod algorithm having a much better performance in terms of the mean task completion time. As of future work, the load balancing (scheduling) problem for the affinity scheme where there are arbitrary number of data localities is an interesting problem which can be studied for investigating the heavy-traffic optimality of the Balanced-Pandas algorithm. Note that the JSQ-MaxWeight is known not to be heavy-traffic optimal for even a system with two levels of data locality, but the Balanced-Pandas algorithm is proven to be delay optimal for a system with three levels of data locality. Extending the delay optimality proof of the Balanced-Pandas algorithm to a system with $N$ levels of data locality, or even proving it for a special traffic load can be of great interest since the affinity scheduling problem is open for nearly three decades. Another interesting future work is to use fluid limit analysis to find how much we may lose in performance when using the Balanced-Pandas-Pod algorithm in high loads very close to the boundary of the capacity region. One can also use water-filling mixed with the power of the $d$ choices algorithm proposed in \cite{ying2017power} to investigate the performance of the Balanced-Pandas-Pod algorithm in a job based system. The load balancing algorithms proposed and discussed in this paper can be used in various applications from supermarkets, signal transmission over different channels, traffic scheduling, and assigning patients to doctors to load balancing for data centers.

\newpage

\bibliographystyle{IEEEtran}
\bibliography{IEEEabrv,qualref}

\end{document}